\documentclass[letterpaper,twoside,twocolumn,english,aps,prl,showpacs]{revtex4}
\usepackage{ae,aecompl}
\usepackage[T1]{fontenc}
\usepackage[latin9]{inputenc}
\usepackage{amsmath}
\usepackage{graphicx}
\usepackage{amssymb}

\makeatletter

\newcommand{\noun}[1]{\textsc{#1}}

\@ifundefined{textcolor}{}
{%
 \definecolor{BLACK}{gray}{0}
 \definecolor{WHITE}{gray}{1}
 \definecolor{RED}{rgb}{1,0,0}
 \definecolor{GREEN}{rgb}{0,1,0}
 \definecolor{BLUE}{rgb}{0,0,1}
 \definecolor{CYAN}{cmyk}{1,0,0,0}
 \definecolor{MAGENTA}{cmyk}{0,1,0,0}
 \definecolor{YELLOW}{cmyk}{0,0,1,0}
 }

\@ifundefined{definecolor}
 {\usepackage{color}}{}
\makeatother

\makeatother

\usepackage{babel}

\begin{document}

\title{The itinerant ferromagnetic phase of the Hubbard model}

\author{Giuseppe Carleo}

\author{Saverio Moroni}

\author{Federico Becca}

\author{Stefano Baroni}

\affiliation{\noun{SISSA} \textendash{} Scuola Internazionale Superiore di Studi
Avanzati \\
 and CNR-IOM \noun{DEMOCRITOS} Simulation Center, Via Bonomea 265
I-34136 Trieste, Italy}

\begin{abstract}
Using a newly developed quantum Monte Carlo technique, we provide
strong evidence for the stability of a saturated ferromagnetic phase
in the high-density regime of the two-dimensional infinite-$U$ Hubbard
model. By decreasing the electron density, a discontinuous transition
to a paramagnetic phase is observed, accompanied by a divergence of
the susceptibility on the paramagnetic side. This behavior, resulting
from a high degeneracy among different spin sectors, is consistent
with an infinite-order phase transition. The remarkable stability
of itinerant ferromagnetism renews the hope to describe this phenomenon
within a purely kinetic mechanism and will facilitate the validation
of experimental quantum simulators with cold atoms loaded in optical
lattices.
\end{abstract}
\maketitle
Ever since classical antiquity, ferromagnetism has attracted the attention
of natural philosopers. ~\cite{Lucretius} A proper understanding
of this phenomenon was only made possible by the advent of quantum
mechanics, from the early interpretations~\cite{Bloch:1929,Stoner:1948}
to its modern realizations in \textsl{quantum simulators} engineered
by means of cold atomic gases.~\cite{Ketterle:2009} In some solids,
such as transitions metals, the spin-independent nature of interactions
has led to conjecture that long-range magnetic order is due to an
itinerant mechanism in which the Coulomb interaction and the Pauli
exclusion principle play a fundamental role. The single-band Hubbard
model, possibly the simplest and most studied lattice model of correlated
electrons, was first thought to encompass a minimal description of
itinerant ferromagnetism.~\cite{Hubbard:1963} Recent experiments
on ultra-cold atoms have shown that a gas of fermions may become ferromagnetic
because of repulsive interactions.~\cite{Ketterle:2009} This important
result and subsequent numerical calculations in the continuum~\cite{Giorgini:2010}
suggested that this phenomenon is universal and independent upon the
details of the interaction, thus renewing the interest in the Hubbard
model with a large Coulomb repulsion, $U$. In spite of its simplicity,
exact solutions of the Hubbard model are not available in more than
one spatial dimension, leaving the question of the stability of a
ferromagnetic phase unsolved. One of the very few exact results that
is known is due to Nagaoka,~\cite{Nagaoka:1966} who proved a theorem
stating that, in the infinite-$U$ limit, a single hole stabilizes
a fully polarized ground state. Following this pioneering work, much
effort has been devoted to study the fully-polarized state for finite
hole densities.~\cite{Doucot:1989,Shastry:1990,Wurth:1995,Gebhard:1991,Becca:2001,Putikka:1992,Park:2008}.
However, a general consensus on the stability of ferromagnetic phases
is still lacking.

In this Letter we present new results for the infinite-$U$ Hubbard
model, based on accurate fermionic quantum Monte Carlo (QMC) simulations,
which indicate that at high electron density the Nagaoka state is
stable not only with respect to the paramagnetic phase, but also with
respect to other previously proposed partially polarized states.~\cite{Becca:2001}
A non-trivial transition to a paramagnetic phase is observed upon
decreasing the electron density. Near the transition this phase is
characterized by highly degenerate states with different values of
the total spin, thus indicating a divergence of the magnetic susceptibility,
consistent with an infinite-order phase transition.~\cite{Benguigui:1977}

The QMC simulation of systems of interacting electrons is beset by
the antisymmetry of the ground-state wave-function which, at variance
with bosons, prevents it from being treated as the stationary distribution
of a diffusion process. The main attempt to cope with the ensuing
difficulties is the so-called \textsl{fixed-node} (FN) approximation,
which, for lattice models, amounts to defining an effective Hamiltonian
whose ground state energy is a variational upper bound to the exact
energy.~\cite{Saarloos:1995} If complemented by an accurate variational
ansatz for the wave-function, the FN method provides a method to study
the properties of large fermionic systems making possible reliable
extrapolations to the thermodynamic limit. Unfortunately, the nature
of the approximation does not allow for an estimate of the residual
error, which not rarely can lead to biased results. However, the infinite-$U$
Hubbard model belongs to an interesting class of Hamiltonians whose
eigenstates of fermionic symmetry are sufficiently close in energy
to the bosonic ground state, to allow them to be treated on an equal
footing; for this class of Hamiltonians we propose a strategy to overcome
the sign problem via the dissection of the excitation spectrum of
the corresponding bosonic auxiliary problem, providing an essentially
{\em unbiased} scheme for medium-size fermionic systems.

\textit{Fermionic-correlation method.} The spectrum of a Hamiltonian
of identical particles, $\mathcal{H}$, can be classified according
to the irreducible representations of the symmetric (permutation)
group. The Pauli principle states that only totally antisymmetric
states are physically allowed for fermions, but mathematical states
of any symmetry can also be considered. In particular, the (unphysical)
state of lowest energy is in general totally symmetric, so that the
fermionic ground state can be formally considered as an excited state
of a bosonic system. As such, it can be studied via excited-state
techniques, provided the Bose-Fermi gap is not too large with respect
to the physical gap in the fermionic sector of the spectrum. Let $|\Psi_{b}^{0}\rangle$
be the bosonic ground state of the system and $\mathcal{A}$ an arbitrary
observable. A recent extension of the reptation QMC method \cite{Baroni:1999}
to lattice models \cite{Carleo:2010} allows for an efficient and
unbiased evaluation of imaginary-time $\tau=it$ correlation functions,
$\mathcal{C}_{\mathcal{A}}(\tau)=\langle\Psi_{b}^{0}|\mathcal{A}^{\dagger}(\tau)\mathcal{A}|\Psi_{b}^{0}\rangle/\langle\Psi_{b}^{0}|\Psi_{b}^{0}\rangle$,
where $\mathcal{A}(\tau)=\mathrm{e}^{\mathcal{H}\tau}\mathcal{A}\mathrm{e}^{-\mathcal{H}\tau}$
is the Heisenberg representation of $\mathcal{A}$.

The connection of such correlation functions with the excited states
$|\Psi^{k}\rangle$ of $\mathcal{H}$ is obtained by considering the
Lehman spectral representation, \begin{equation}
\mathcal{C}_{\mathcal{A}}(\tau)=\frac{\sum_{k}|\langle\Psi_{b}^{0}|\mathcal{A}|\Psi^{k}\rangle|^{2}e^{-\Delta_{k}\tau}}{\langle\Psi_{b}^{0}|\Psi_{b}^{0}\rangle},\label{eq:cabtau}\end{equation}
 where $\Delta_{k}=E^{k}-E_{b}^{0}$ are excitation energies with
respect to the bosonic ground state. Selection rules act in such a
way as to exclude from Eq. \eqref{eq:cabtau} those excited states
whose symmetry is different from that of the state $\mathcal{A}|\Psi_{b}^{0}\rangle$.
In particular, if $\mathcal{A}$ is chosen to be totally antisymmetric
with respect to permutations, only fermionic (ground and excited)
states would contribute to $\mathcal{C}_{\mathcal{A}}(\tau)$. For
example, if $\mathcal{A}$ is the local operator whose coordinate
representation is the ratio between the fermionic and bosonic ground-state
wave-functions ($\mathcal{A}_{f}(n)=\langle n|\Psi_{f}^{0}\rangle/\langle n|\Phi_{b}^{0}\rangle$,
where $|n\rangle$ denotes the many-body lattice configuration), the
correlation function $\mathcal{C}_{\mathcal{A}}(\tau)$ would be proportional
to the single exponential $\mathrm{e}^{-\Delta_{0}\tau}$.

In practice, neither the bosonic nor the fermionic ground state are
known exactly and only variational approximations to them are available,
which we denote here by $|\Phi_{b}\rangle$ and $|\Phi_{f}\rangle$,
respectively. Correspondingly, the antisymmetric observable is defined
as $\mathcal{A}_{f}(n)=\langle n|\Phi_{f}\rangle/\langle n|\Phi_{b}\rangle$.
In this way, the leading coefficient of the expansion is given by
${\langle\Psi_{b}^{0}|\mathcal{A}_{f}|\Psi_{f}^{0}\rangle\simeq\langle\Phi_{f}|\Psi_{f}^{0}\rangle}$
and can be systematically maximized improving the quality of the variational
states. The energy of the fermionic ground state can be then extracted
either directly by noticing that $E_{f}^{0}=E_{b}^{0}-\lim_{\tau\rightarrow\infty}\left[\partial_{\tau}\log\mathcal{C}_{\mathcal{A}}(\tau)\right]$
or, indirectly, by fitting the exponential decay of the correlation
function of Eq.~(\ref{eq:cabtau}) and extracting the smallest energy
gap. In order for this procedure to make any sense, it is necessary
that the (unphysical) Bose-Fermi gap is not too large with respect
to the physical excitation energies in the fermionic sector of the
spectrum. If this condition is not met, the anti-symmetric correlation
function gets effectively extinguished before the selection of the
fermionic ground state from its excitation background is attained
by imaginary-time evolution. This condition is actually verified for
infinite-$U$ fermionic Hubbard models of moderate size, where the
properties of the system are not too dissimilar from those of a system
of hard-core bosons. The condition of a small fermion-boson gap is
also met in other interesting systems, where the effects of statistics
on the total energy are overwhelmed by the effects of correlations,
such as the low-density electron gas, liquid $^{3}$He, quasi-unidimensional
systems and mixtures of bosons and fermions.

We remark that even in the most favorable cases all these considerations
can only hold for systems of moderate size because the Fermi-Bose
gap is an extensive property, whereas the physical gap in the Fermi
sector of the spectrum is intensive, being determined by quasi-particle
effects. This fermionic-correlation method is related to the \emph{transient
estimate} (TE) method for the fermionic ground state,~\cite{Schmidt:1984}
or its generalization for a few excitations.~\cite{Bernu:1988} However,
TE works with \emph{ratios} of decaying correlation functions, thereby
reducing the signal/noise ratio, and typically uses sub-optimal bosonic
guiding functions, with increased fluctuations in the weights of the
random walks.

The calculation of spectra from imaginary-time correlation functions
is in general an ill-posed problem. In practice, however, sharp peaks
with strong spectral weight can be reliably extracted if the correlation
function is known with good statistical precision for sufficiently
long times.~\cite{Baroni:1999,Moroni:2003} In the present work this
condition is met even for systems of several tens particles, due to
the relatively small gap between the fermionic and bosonic ground
states, as well as to the good quality of the variational states.

\begin{figure}
\includegraphics[width=0.95\columnwidth]{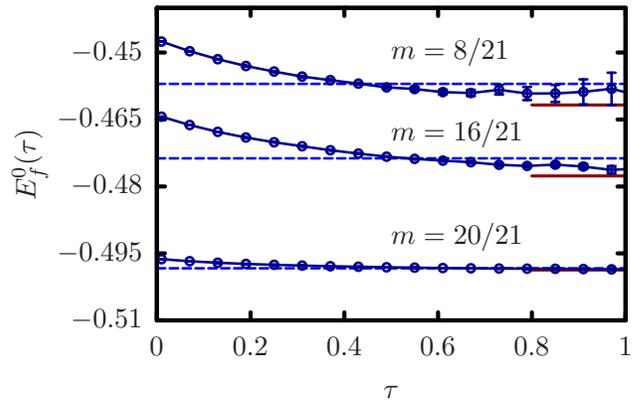} \caption{\label{fig:Energy-Corr} Energy $E_{f}^{0}(\tau)=E_{b}^{0}-\partial_{\tau}\log\mathcal{C}_{\mathcal{A}}(\tau)$
as a function of the imaginary-time $\tau$ for $L=50$ and $N=42$
electrons and different magnetizations. The dashed horizontal lines
are FN energies, while the solid lines are the energies as obtained
fitting the imaginary-time correlations.}

\end{figure}

\textit{The model.} The Hamiltonian of the infinite-$U$ Hubbard model
reads: \begin{equation}
\mathcal{H}_{f}=-t\sum_{\langle i,j\rangle,\sigma}\mathcal{P}_{G}c_{i,\sigma}^{\dagger}c_{j,\sigma}\mathcal{P}_{G}+h.c.,\label{eq:hamilt}\end{equation}
 where $c_{i,\sigma}^{\dagger}$ ($c_{i,\sigma}$) creates (destroys)
an electron on site $i$ with spin $\sigma$; $\langle i,j\rangle$
denotes nearest-neighbor site pairs and the Gutzwiller projector $\mathcal{P}_{G}$
forbids double-occupancy. In the following, we will consider a square
lattice and take $t=1$ as the energy scale. The total number of sites
will be denoted by $L$ and the number of electrons by $N$. In the
following, we present the results for different magnetizations $m=(n_{\uparrow}-n_{\downarrow})/(n_{\uparrow}+n_{\downarrow})$
and densities $n=n_{\uparrow}+n_{\downarrow}$.

Relatively simple variational wave functions have been constructed,~\cite{Shastry:1990,Wurth:1995}
by flipping one (say up) spin with respect to the saturated ferromagnetic
state. The flip of the spin leads to a gain of kinetic energy for
the down spin, but also a loss in the spin-up kinetic energy (since
the motion of spin up electrons is restricted by the necessity of
avoiding double occupancy). Here, we consider \begin{equation}
\langle n|\Phi_{f}\rangle=\mathcal{J}^{f}(n)\times\text{Det}\left\{ \phi_{k}(R_{j}^{\uparrow})\right\} \times\text{Det}\left\{ \phi_{k}(R_{j}^{\downarrow})\right\} ,\label{eq:phifermi}\end{equation}
 where the Jastrow factor $\mathcal{J}^{f}(n)=\exp\left[\sum_{i,j}V_{ij}^{f}n_{i}n_{j}\right]$
multiplies two Slater determinants that are constructed by applying
backflow correlations to single-particle orbitals for up and down
spins.~\cite{Tocchio:2008} The correlated orbitals are defined by
$\phi_{k}(R_{j}^{\sigma})=\phi_{k}^{0}(R_{j}^{\sigma})+b_{k}\sum_{R_{l},\sigma^{\prime}}\phi_{k}^{0}(R_{l}^{\sigma^{\prime}})$,
where $b_{k}$ are orbital-dependent backflow parameters, $\phi_{i}^{0}(R_{j}^{\sigma})$
are plane waves, and the sum includes all nearest neighbors of the
$j$-th particle, thus preserving the spin rotational invariance.
The proposed backflow wave function~(\ref{eq:phifermi}) encodes
the effect of correlation on the deformation of the free-orbitals
nodal structure and consistently catches much of the physics of previous
treatments,~\cite{Shastry:1990,Wurth:1995,Gebhard:1991} while leaving
room for a systematic improvement with the QMC methods.

\begin{figure}[t]
 \includegraphics[width=0.9\columnwidth]{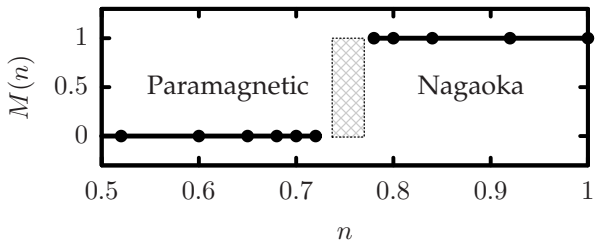} \caption{\label{fig:Magn} Ground-state magnetization of the infinite-$U$
Hubbard model on the square lattice. The shaded area represents a
small region of uncertain attribution due to the effect of the residual
Monte Carlo error.}

\end{figure}

\begin{figure}[b]
 \includegraphics[width=0.82\columnwidth]{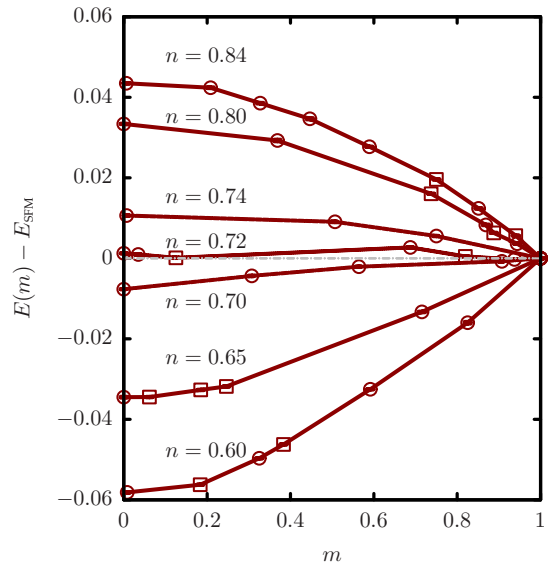} \caption{\label{fig:Energy-V0} Difference between the energy per site of different
magnetizations and the one of the saturated ferromagnet as a function
of the density $n$. The cases with $L=200$ (squares) and $L=400$
(circles) are reported; lines connecting points are a guide for the
eye.}

\end{figure}

The bosonic counterpart of the model studied is a purely kinetic hard-core
bosons Hamiltonian, where the fermionic operators are substituted
by bosonic ones. Our QMC method is particularly suitable to study
the high-density region, namely few holes close to full filling ($N=L$),
where the boson-fermion gap is very small and increases upon decreasing
the density.~\cite{Becca:2000} The bosonic trial state is given
by a Jastrow wave function $\langle n|\Phi_{b}\rangle=\mathcal{J}^{b}(n)$,
which is similar to the fermionic one (but with different parameters
$V_{i,j}^{b}$) and represents an excellent ansatz for the bosonic
ground state.~\cite{Capello:2007} In all cases, the variational
parameters are fully optimized minimizing the variational energy with
the method of Ref.~\onlinecite{Sorella:2005}.

\textit{Results.} The fermionic correlation technique remains efficient
up to relatively large system sizes (i.e., $L=50\div100$) and allows
us to reach numerical results, which are exact within statistical
accuracy. In Fig.~\ref{fig:Energy-Corr}, we report our results for
$L=50$ and $N=42$ electrons, for different values of the magnetization,
$m$. In addition, we also report the results based upon the FN approach.
The possibility to obtain numerically exact results on rather large
systems allows us to assess the accuracy of the FN method that can
be extended to much larger sizes (i.e., $L\lesssim1000$), without
any numerical instability. Thanks to backflow correlations, we get
a considerable improvement upon the standard plane waves that has
been used in Ref.~\onlinecite{Becca:2001}. There is a small difference
between the FN results and the energies obtained by the imaginary-time
correlations, indicating a very small residual FN error, namely $\Delta E/t\lesssim0.01$. 

In Fig.~\ref{fig:Magn}, we report the overall phase diagram obtained
by considering large-scale FN calculations. A saturated ferromagnetic
phase is stable for $n\gtrsim0.75$, while for smaller densities a
paramagnetic ground state is found. The narrow shaded region denotes
the incertitude due to the residual numerical error, which can be
estimated by comparing the FN energies with the \textsl{exact ones}
(obtained from the fermionic correlations) on smaller clusters, see
Fig.~\ref{fig:Energy-Corr}. This direct comparison puts us on secure
grounds as concerns the robustness of the dependence of the ground-state
magnetization on the electron density.

In Fig.~\ref{fig:Energy-V0} we display the dependence of the ground-state
energy upon magnetization, for different values of the electron density.
The remarkable feature emerging from this figure is the strong flattening
of the energy as a function of the magnetization (i.e., the spin)
close to the transition between the fully polarized ferromagnet and
the paramagnetic state. Indeed, at low and high densities the energy
has a monotonic behavior as a function of the magnetization $m$.
At low density a clear minimum exists at $m=0$, typical of a paramagnetic
phase, where the curvature of the energy-versus-magnetization curve
witnesses to a finite spin susceptibility. On the other hand, in the
high-density ferromagnetic phase, $E(m)$ displays a well defined
minimum for $m=1$. By approaching the transition, $E(m)$ becomes
flatter and flatter, suggesting that the susceptibility may diverge
at the critical point. Although we cannot exclude a tiny region with
a finite but non-saturated magnetization, these results would suggest
that the paramagnetic-to-ferromagnetic transition is not due to a
simple level crossing, namely to the creation of a local minimum in
$E(m)$ at $m=1$ that eventually prevails over the paramagnetic one,
but rather to the progressive flattening of the \textit{whole} $E(m)$
curve. 

Our scenario is compatible with an infinite-order phase transition,
which, in general, is described by $E(m)=(g-g_{c})m^{2}+bm^{2r}$,
where $r\to\infty$; a phase transition is obtained by varying the
order parameter $g$ (in our case the electron density) across its
critical value $g_{c}$. The critical exponent of the order parameter
is $\beta=1/(2r-2)$, generating a jump from zero to the saturation
value for $r\to\infty$. Moreover, the susceptibility $\chi\sim A_{\pm}/|g-g_{c}|^{\gamma}$
has an exponent $\gamma=1$ independent of $r$, with an amplitude
ratio $A_{-}/A_{+}$ that vanishes for $r\to\infty$.~\cite{Benguigui:1977}
Even though the order parameter shows a finite jump, like in ordinary
first-order phase transitions, there is no hysteresis. We have indeed
verified that the ground-state energy is a convex function of the
electron density, implying a finite compressibility in the neighborhood
of the ferromagnetic-paramagnetic transition. This picture implies
that spin-flip excitations over the fully polarized state are non-interacting
at the transition point. In fact, we find that, at small distances,
the minority spins repel each other, whereas at large distances they
do not interact. In the variational wave function, this fact generates
a sizable repulsive short-range Jastrow factor, while at long range
the $V_{i,j}^{f}$ \textsl{pseudopotential} vanishes.

\textit{Conclusions.} In this Letter we have analyzed with high accuracy
the magnetic phase diagram of the fermionic Hubbard model on the square
lattice in the limit of infinite on-site repulsion $U$. By the combination
of different QMC methods, we are able to give a very precise determination
of the transition between the ferromagnetic and the paramagnetic states.
Interestingly, all spin excitations become essentially gapless at
the transition, possibly indicating that the transition is of infinite
order.~\cite{Benguigui:1977} Compared to previous calculations,
this is the first time that such a behavior is observed. Indeed, given
the extreme difficulty to treat this highly-correlated system, most
of the theoretical efforts were limited to study very high densities
or a single spin flip.~\cite{Doucot:1989,Shastry:1990,Wurth:1995,Gebhard:1991}
%Other kind of approaches are based upon systematic expansions around infinite
%temperature or dimensionality~\cite{Putikka:1992,Park:2008} and may miss to
%capture specific aspects of the two-dimensional ground state.

Our results pave the way to a better understanding of itinerant ferromagnetic
phenomena in both traditional condensed matter systems and in recent
and forthcoming realizations of such phases in cold atomic gases.
Indeed, the recent achievements for the realization of interacting
fermionic systems trapped in optical lattices~\cite{Esslinger:2008}
will most likely lead to experimentally probe the strongly-correlated
regime of the Hubbard model at sufficiently low temperatures. Finally,
the generality of the numerical methods introduced in this Letter
will also offer new insights in other strongly correlated fermionic
systems where currently available analytical and numerical treatments
may fail to offer a quantitative or even qualitative account of the
relevant physical properties.

We thank S. Sorella and A. Parola for many discussions. We acknowledge
support from COFIN07 and CASPUR, through the standard HPC Grant 2010.

\bibliographystyle{apsrev} 

\end{document}